\documentclass{PoS}
\title{Large Field Inflation/Quintessence  and the Refined Swampland Distance Conjecture}

\ShortTitle{Large Field Inflation/Quintessence and the RSDC}

\author{\speaker{Ralph Blumenhagen} \\
        Max-Planck-Institut f\"ur Physik (Werner-Heisenberg-Institut), \\ 
   F\"ohringer Ring 6,  80805 M\"unchen, Germany \\
        E-mail: \email{blumenha@mpp.mpg.de}}

\abstract{Attempts to construct string derived effective field theory
  models realizing  large field inflation are plagued
   by control issues. Targeted at  a broader audience,
     in this article we review
    recent progress in isolating the underlying conceptual reasons for this failure.
   Special emphasis is given to models of axion monodromy inflation and
   their relation to the Swampland Distance Conjecture. This  discriminates 
   effective actions that admit a UV completion, the landscape,  from those that do not,
   the swampland. Since they are conceptually very similar, we also comment on 
   implied challenges for axionic quintessence models. 
}

\FullConference{Corfu Summer Institute 2017 ``Workshop on Testing Fundamental Physics Principles'',
22 - 28 September, 2017,Corfu, Greece,\\ Preprint: MPP-2018-67}

\usepackage{amsmath}
\usepackage{amsfonts}
\usepackage{amssymb}
\usepackage{graphicx}
\usepackage{tikz} 
\usetikzlibrary{decorations.pathreplacing}

\newcommand{\eq}[1]{\begin{equation}
                     \begin{split} #1 \end{split}
                     \end{equation}}

\newcommand{\ov}{\overline}

\allowdisplaybreaks[2]
\numberwithin{equation}{section}

\begin{document}


\section{Introduction}

It is widely accepted that string theory features  a landscape of consistent 
solutions to its equations of motion. Historically, such a landscape has emerged
in two disguises. First, by studying and partially classifying  
string compactifications to four dimensions with N=1 supersymmetry.
Examples are covariant lattice constructions \cite{Lerche:1986cx}, Gepner type constructions \cite{Schellekens:1989wx,Blumenhagen:1996vu},
Calabi-Yau compactifications with line-bundles \cite{Lebedev:2006kn,Anderson:2011ns}
or intersecting D6-brane models \cite{Gmeiner:2005vz,Ibanez:2007rs}.  It became clear that instead of a small
set of such  four-dimensional solutions, there exist a whole plethora of them. 
For instance, in case of the covariant lattice construction estimates
of the order $O(10^{1500})$ were given.  

Each of these models comes with a number
of still massless scalar and pseudo-scalar fields, so that they do not correspond to isolated minima,
but live in a higher dimensional moduli space. The pseudo-scalar fields are
also called axions and being CP-odd and obeying a perturbative shift
symmetry, they couple very differently than the scalars in the effective theory. 
The latter are also called saxions and they are dangerous
for the low-energy phenomenology as they can give rise to (yet undetected)
fifth forces and spoil the history of the early universe. Indeed, if the 
mass of a scalar is below $\sim 30\,$TeV, its gravitational decay
will lead to a reheating temperature below $T_{\rm BBN}\sim 1\,$MeV, so that 
the moduli would decay after nucleosynthesis. This spoils the thermal history of the
early universe and is called the cosmological moduli problem. On the
contrary, some of the  axions can stay  very light with masses in the
sub-meV regime  and  contribute to dark matter.

Therefore, eventually the scalar  moduli should better be absent or should be
stabilized by some mechanism, respectively. There are essentially
two working proposals how this could happen. Either non-perturbative
effects  generate an exponentially suppressed potential for the moduli
or they receive already a tree-level potential by turning on more general
background fields, so called fluxes. Concerning the latter mechanism, it was shown that 
type IIB orientifolds compactified on Calabi-Yau threefolds show a discretuum
of Minkowski minima \cite{Ashok:2003gk} in the complex structure and the axio-dilaton moduli. Here, except
for the K\"ahler moduli, generally all  minima are isolated and simple estimates gave the
(in-)famous number of $10^{500}$ vacua. This is the second way a landscape arose
in the history of string theory and it  gave rise to the
idea of a multiverse that possibly ameliorates  some fine tuning
problems, like the problem of the smallness of the cosmological
constant in our universe.

In view of this exponentially large number it was questioned whether string theory is able
to make any concrete predictions at all, as it appears that almost everything goes.
However, that this is not the case is clear at least for each type  of concrete string construction.
For instance, there are strong tadpole constraints to hold that give an upper bound
on the rank for the gauge group. Moreover, the general experience is that whenever
one hunts for a concrete model with certain properties, the stringy constraints
are severe and finding a model it is not trivial at all. For instance,  isolating
a supersymmetric Madrid-type intersecting D6-brane model with three generations
in the Standard Model gauge group and massless hypercharge was challenging and
only one in a billion models satisfied these rough discrete
constraints \cite{Gmeiner:2005vz}.
After all, effective four-dimensional models that can be realized in string theory
are guaranteed to have a consistent UV completion. 

In the last years the opposite question was under scrutiny, namely whether string
theory can in principle be falsified. For that purpose, one takes the
concrete model building obstacles seriously and tries to formulate well motivated 
conjectures that distinguish effective field theories with a 
UV completion (the landscape) from those which do not admit it (the swampland).
This logic can in principle lead to statements like, ''the physical property A is in the swampland
of effective field theories derived from string theory''. In this case, if property A is experimentally measured in our universe, then
it cannot  be described by an effective field theory arising from string theory.
Logically is not yet clear whether string theory is really falsified or our logic about effective field theories does not apply. In any case, by better understanding properties that lie in the swampland
of effective field theories derived from string theory, we learn a lot on the underlying conceptual  principles of  quantum gravity and string theory, respectively.

One potential candidate for such an enterprise arises in cosmology, namely  large field inflation.
Interestingly triggered by the, at first  misinterpreted, BICEP2
measurement of the cosmic microwave background (CMB), 
during the last four years the community made a big effort  in realizing large field inflation
in string theory. Such models feature a tensor-to-scalar ratio in a regime
that can be measured with current or planed experiments.
As we will briefly review, there exist various proposals for getting
a controllable model of string inflation. 
They all exploit the shift symmetry of axions and can be 
divided into  two classes. In the first class, the shift symmetry of the axion is broken
to a discrete one by non-perturbative effects. In the second class, a branch structure
arises, where each branch features a polynomial potential for the axion and the shift symmetry
is still present but shifts also the branch for the potential. In other words, by a choice
of branch the shift symmetry is spontaneously broken and the axion is not periodic any more.
These models go under the name of axion monodromy inflation \cite{Silverstein:2008sg,McAllister:2008hb}
(see \cite{Dvali:2005an,Kaloper:2008fb,Kaloper:2011jz} for field theory analogues).
The question arises is whether the large (trans-Planckian) field regime of such models is under
control in the string derived effective field theory.

In this talk, for a broader audience I will review  recent attempts to clarify this issue. Methodologically, 
we will employ  a combination of concrete string model building attempts
and abstracted swampland conjectures on UV complete quantum gravity derived effective field theories.
In section 2, a brief reminder of some basic notions of large single field inflation and also of
quintessence models is presented. In section 3, I review string theory proposals for
the realization of axion inflation via non-perturbative effects and how they are bound to fail, 
if the Weak Gravity Conjecture (WGC) holds.
A second swampland conjecture on the validity of quantum gravity derived effective field theories
is introduces in section 4.  This is called 
the Refined  Swampland Distance Conjecture (RSDC) and quantitative evidence for it is summarized,
that arose from a detailed study of distances in concrete  Calabi-Yau moduli spaces.
In section 5, models of axion monodromy inflation are introduced and after providing a concrete example
it is argued that  such models are challenged by an axionic extension of
the  RSDC.
Section 6 contains a few  comments on how also models of axionic quintessence face
the same challenges. In section 7 we summarize a few more swampland conjectures
that have recently been proposed in the literature, the most dramatic
one stating that meta-stable de Sitter minima might also be in the swampland.

\newpage 

\section{Preliminaries}

\begin{figure}[ht]
\begin{center} 
\includegraphics[width=0.37\textwidth]{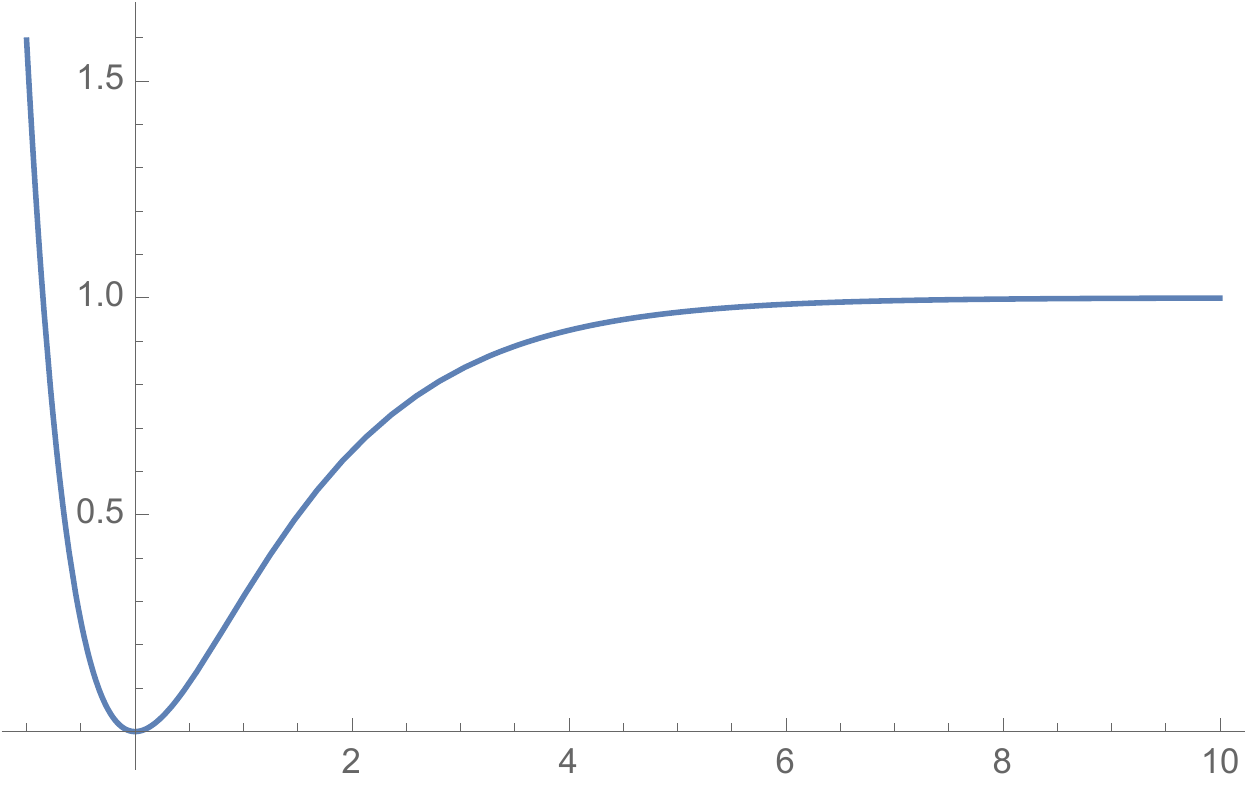}
\begin{picture}(0,0)
  \put(0,3){$\Theta$}
   \put(-147,103){$V$}
  \put(-60,62){$\leftarrow\; \bullet$}
   \put(-130,36){$\swarrow$}
  \put(-118,45){$\bullet$}
  \put(-118,20){$\leftarrow\quad\   \ \Delta\Theta\quad\   \ \rightarrow$}
\end{picture}
\hspace*{15pt}
\end{center} 
\vspace{-5pt}
\label{fig_A}
\end{figure}

Not only provides it a natural  theoretical explanation  of the homogeneity and flatness
properties of the observable  universe, but 
there is now also mounting detailed experimental evidence for the existence of an inflationary epoch in the 
early universe. In particular, the temperature fluctuations in the CMB
are predicted by such a scenario in an impressive manner.
The combined  PLANCK 2015 and  BICEP2/Keck Array  results \cite{Array:2015xqh} confirm that these fluctuations are almost scale
invariant and Gaussian. Moreover, these measurements provide an upper bound to the  tensor-to-scalar ratio.
For the corresponding  parameters they find 
\begin{itemize}
\item{tensor-to-scalar ratio: $r<0.07$}
\item{spectral index: $n_s=0.9667\pm 0.004$ and its running
    $\alpha_{s}=-0.002\pm 0.013$}
\item{amplitude of the scalar power spectrum ${\cal P}=(2.142\pm
    0.049)\cdot 10^{-9}$ \,.}
\end{itemize}
These findings are consistent with a simple model of inflation, namely that of
a single slowly rolling scalar field as shown in figure \ref{fig_A}.

\vspace{0.2cm}
\vspace{0.2cm}

\noindent
Here $\Delta\Theta$ denotes the distance in field space along which the 
inflaton rolls, i.e. where the slow-roll conditions 
\eq{
          \varepsilon={M_{\rm pl}^2\over 2} \left({ V'\over V}\right)^2\ll 1\,, \qquad
            \eta={M_{\rm pl}^2} \left({ V''\over V}\right)\ll 1
}
are satisfied. 
The measured parameter can be expressed in terms of the slow-roll parameters as
\eq{
      n_s= 1+2\eta-6\varepsilon\,,\qquad r=16\varepsilon\,,\qquad
          {\cal P}={H_{\rm inf}^2\over 8\pi^2  \varepsilon  M^2_{\rm pl}}
}
where $H_{\rm inf}$ denotes the Hubble constant during inflation.
Due to  the Lyth bound \cite{Lyth:1996im}
\eq{
        {\Delta \Theta\over M_{\rm pl}}>O(1)\sqrt{r\over 0.01}
}
the distance in field space traversed during inflation is directly related to the tensor-to-scalar ratio.
Thus,  if $r$ is detected by the current experiments then theoretically 
a model of  large field inflation will be favored, i.e.
$\Delta \Theta>M_{\rm pl}$.
Moreover, in this case  the Hubble constant during inflation
comes out as  $H_{\rm inf}\sim 10^{14}\,{\rm GeV}$. Via $V_{\rm inf}=3M^2_{\rm pl} H_{\rm inf}^2$
one can infer  that the inflationary mass scale is  of
the order of the GUT scale
\eq{   M_{\rm inf}=(V_{\rm inf})^{1\over 4}\sim \left({r\over
      0.1}\right)^{1\over 4} \times 1.8\cdot 10^{16}\,{\rm GeV} \,.
}
The mass of the inflaton is given by $M_\Theta^2=3\eta M^2_{\rm pl} H_{\rm inf}^2$
and comes out as $M_{\Theta}\sim 10^{13}\,{\rm GeV}$.
Thus, all mass scales are close to the GUT scale and only one order of magnitude
apart.

Recalling that inflation describes a period of a slowly rolling scalar field  in the early
universe
giving rise to  an effective  positive cosmological constant, it is tempting to speculate that also the very tiny positive cosmological constant in 
our present universe can be originating via a similar mechanism.
This proposal goes under the name of quintessence \cite{Ratra:1987rm} and conceptually  it is very similar
to inflation, the main difference being that the mass scales of quintessence are much tinier. 
The Hubble parameter today is of the order $H\sim 10^{-33}\,$ eV and the measured cosmological
constant is
\eq{
                 V_{\rm quint}=10^{-120} \, M_{\rm pl}^4 \sim \Big( 10^{-3}\,{\rm eV} \Big)^4\,.
}
Note that quite non-trivial,  these two scales are consistent with the slow-roll expectation 
\eq{
V_{\rm quint}\sim M^2_{\rm pl} \, H^2\,.
}
The mass of the quintessence field is then of the order
of the Hubble scale, thus very tiny with respect to  e.g. the heavy moduli masses that due
to the cosmological moduli problem must be larger than $30\,$TeV. The equation of state is
$p=w\rho$ where the parameter $w$ has been measured by the Planck
satellite $w=-1.006\pm 0.045$. In the slow-roll regime, this is related to the
slow-roll  parameter $\varepsilon$ as 
\eq{
\label{wstate}
                     w\sim {\varepsilon/3-1\over \varepsilon/3+1}\,.
}

\begin{figure}[ht]
\begin{center} 
\includegraphics[width=0.37\textwidth]{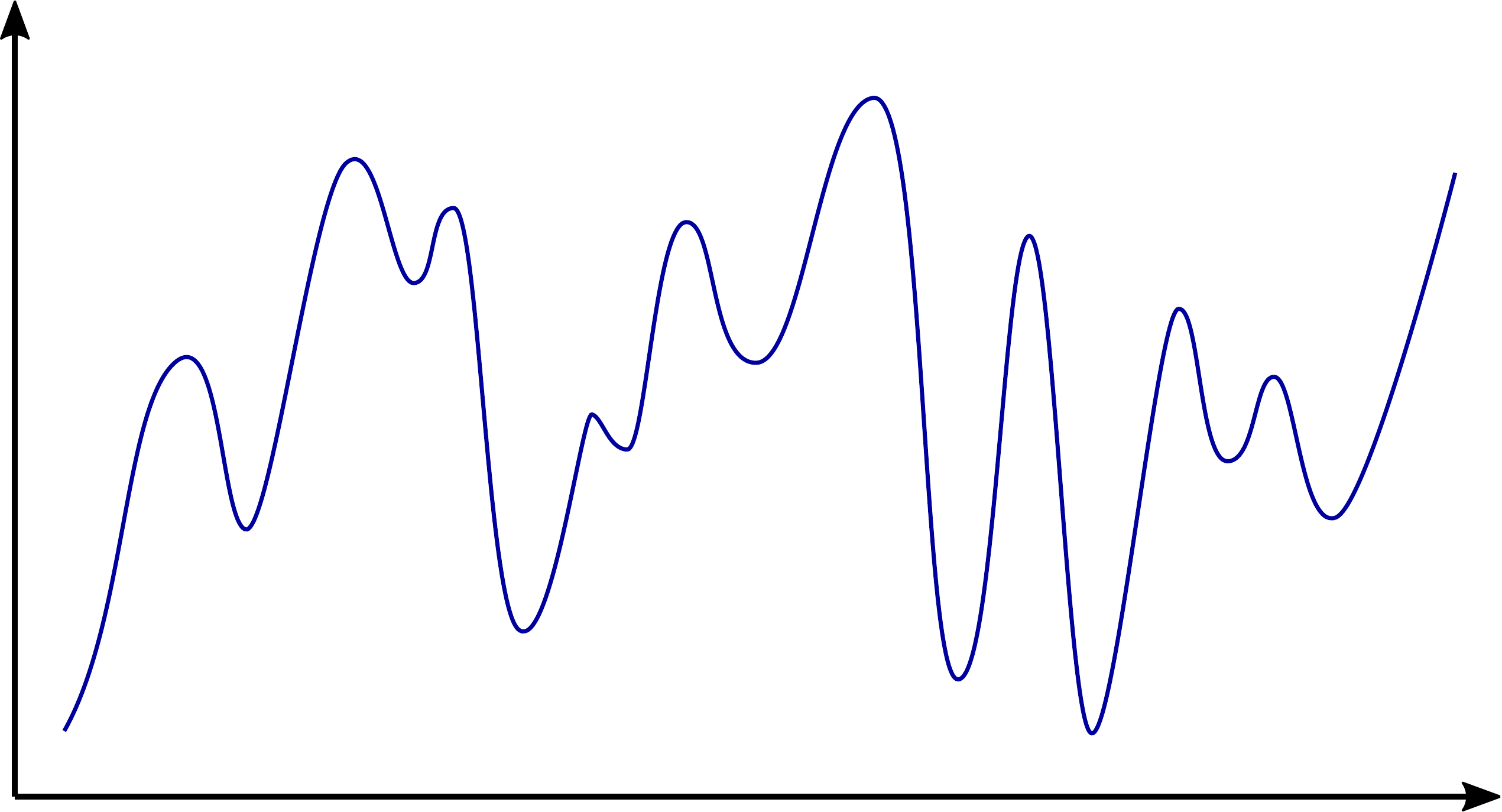}
\begin{picture}(0,0)
  \put(0,0){$\Theta$}
  \put(-163,88){$V$}
  \put(-51,-14){$M_{\rm pl}$}
  \put(-48,-1){$\vert$}
\end{picture}
\end{center} 
\vspace{-5pt}
\label{fig_B}
\end{figure}

The main issue with such slow-rolling  models of inflation (or quintessence) is that they are very sensitive
to corrections to the scalar potential of the inflaton. In particular, one expects that
the effective field theory model has to be embedded into a UV complete theory of
quantum gravity. In such a theory, like string theory, one expects corrections
by Planck-suppressed operators of the form $(\Theta/M_{\rm pl})^n$. However,
for large field inflation, such corrections are substantial and will 
change the potential in an uncontrolled way, as shown in figure \ref{fig_B}.
 
\vspace{0.2cm}

\vspace{0.1cm}
\noindent
For models of quintessence, one has to control the Planck suppressed 
corrections to the mass of the scalar up to order $n=6$ in  
\eq{
                  \Delta M^2 \sim M^2_{\rm susy}\Big( { M_{\rm susy}\over M_{\rm pl}}\Big)^n
}
for a supersymmetry breaking scale $M_{\rm susy}=1\,$TeV. Higher
scales of supersymmetry breaking imply the control of even higher orders.

In order to make progress, one really has to discuss these  issues in an understood
UV complete theory of quantum gravity. In particular, one has to study whether one
can identify  mechanisms to control Planck suppressed operators. As usual
such mechanism will be related to symmetries that will be broken in a controlled
way. Natural candidates are the shift symmetry of axions that can be broken
by non-perturbative corrections. Let us discuss how this can in principle be realized
in the context of string theory compactifications.

\section{Natural  inflation and the weak gravity conjecture}

As mentioned in the introduction, to protect the potential from  receiving 
Planck suppressed corrections, one can employ the shift symmetry of axions.
Such four-dimensional axions arise naturally in string compactifications by either directly
resulting from ten-dimensional axion-like fields, like the NS-NS and R-R p-forms.
In addition, at special points in the moduli space of e.g. the complex structure moduli space,
accidental shift symmetries of some of the scalar fields can arise.
For instance, around the large complex structure   point of a Calabi-Yau (CY) moduli space, the K\"ahler potential
takes the form
\eq{
\label{kaehlera}
                K=-\log \Big(i  \kappa^{ijk} (U_i-\ov U_i)(U_j-\ov U_j)(U_k-\ov U_k)+\ldots \Big)
}
where $\kappa^{ijk}$  denote the triple intersection numbers of the  mirror dual CY threefold.
This K\"ahler potential is shift symmetric in the real part of the complex structure 
moduli $U_i$. This shift symmetry is broken by higher order terms, that correspond
to world-sheet instanton corrections on the mirror dual side.

The simplest class of models assumes that, after stabilizing all heavy moduli except the lightest
axion $\theta$, the continuous shift symmetry of the latter is broken
to a discrete one by a non-perturbative correction to 
the potential. Thus, the effective Lagrangian takes the form
\eq{
\label{Lagnat}
              {\cal L}= f^2 (\partial \theta)^2 + \Lambda (1-\cos\theta) \,
}
where $f$ is the axion decay constant and $\Lambda\sim \exp(-S_{\rm inst})$ the size of 
the (leading order) non-perturbative instanton correction. 
Here we have chosen natural units where  $M_{\rm pl}=1$. 
Note that  one has assumed that
$S_{\rm inst}>1$ so that higher order instanton corrections can be neglected.
Moreover, from the string theory point of view,  we have implicitly assumed that moduli stabilization can be achieved such that
at the end one gets a Minkowski minimum at $\theta=0$. Whether this
is possible in string theory is a not yet completely settled issue, as in most concrete cases
AdS-minima seem to be preferred and one needs to rely on the existence
of some sort of uplift mechanism. More on this will be mentioned in section 5. 

However, this is not the point here,
as already the Lagrangian \eqref{Lagnat} has an inherent problem in the  large 
field regime. To see this let us first transform it to a canonically normalized kinetic term.
With $\Theta=f\theta$ we get
\eq{
\label{Lagnatb}
              {\cal L}= (\partial \Theta)^2 + \Lambda \Big(1-\cos\Big({\Theta\over f}\Big)\Big) \,.
}
In the regime $\Theta/f\ll 1 $ this potential reduces to the one of
chaotic inflation $V\sim \Lambda \Theta^2/f^2$,
which  gives rise to large field inflation with $\Delta\Theta\sim10$ and $r\sim 0.2$.
Therefore also $f$ has to be larger than one.
This field theory model is called natural inflation \cite{Freese:1990rb}. 
Now in string theory, the (pseudo-scalar) axion is combined with a
scalar $\phi$ (also called saxion) to a complex field
$T=\theta+i\phi$ and the parameters in \eqref{Lagnat}  are given by
\eq{
\label{axionf}
             f={1\over \phi}\,,\qquad   S_{\rm inst}=\phi
}
up to numerical factors of order one\footnote{The first relation can be seen as a consequence
of a K\"ahler potential of the type \eqref{kaehlera}. Indeed  the
K\"ahler potential $K=-3\log(-i( T-\ov T))$ implies
for the metric $G_{T\ov T}=\partial_T\partial_{\ov T} K={3\over 4\phi^2}$. }. Thus, for this
string derived effective action, one has the relation 
\eq{
\label{wgc_inst}
                                 f\, S_{\rm inst}\sim 1 \,. 
}
As a consequence, the large field regime $\Theta>1$ (presuming $f>1$) requires
that the instanton action is smaller than one, spoiling the validity of the instanton
expansion. Therefore, the large field regime lies outside the regime
of validity of the pre-assumed effective action. The same logic also applies
if one wants to interprete the axion $\Theta$ as a quintessence field.

Of course this is a just a simple example, but it has been conjectured that the
reason behind this failure  is a property of any consistent theory of quantum gravity.
Namely, as argued in \cite{Rudelius:2015xta,Montero:2015ofa,Brown:2015iha} 
the relation \eqref{wgc_inst} is just the generalization of the so-called Weak Gravity Conjecture (WGC)
\cite{ArkaniHamed:2006dz},
to more general p-forms (here 0-form).
The WGC is formulated for a Maxwell theory with gauge coupling $g$ coupled to gravity and it says
that  gravity must be  the weakest force, i.e.  for a
U(1) gauge theory there must exist a particle of mass $m$ and charge
$q$ such that   $m\le g q M_{\rm pl}$.
Arguments in favor of this conjecture have been given in \cite{ArkaniHamed:2006dz}, some of them being:
\begin{itemize}
\item{The WGC guarantees that any non-BPS subextremal  charged black hole with $M\ge Q$ can decay.}
\item{It is satisfied for perturbative states in toroidal heterotic string theory compactifications.}
\item{It has a magnetic version $\Lambda< g M_{\rm pl}$ relating the  cut-off  of the theory.
            This can be derived by requiring that a magnetic monopole is not a black-hole.
                  It  implies that the  $g\to 0$ limit is dramatic so that one cannot 
           simply generate a continuous global symmetry from a gauge
           symmetry\footnote{It is expected that quantum gravity
             theories do not admit continuous global symmetries.}.} 
\end{itemize}

\noindent
Via T-duality it has been argued that there should exist such
a relation for any p-form gauge field \cite{Rudelius:2015xta,Montero:2015ofa,Brown:2015iha}.
For a 0-form with 
\eq{              
  m \to S_{\rm inst}\,\qquad    gq\to 1/f  
}
one precisely arrives at \eqref{wgc_inst} 
\eq{
                                    f S_{\rm inst}\le 1\,
}
where we have set $M_{\rm pl}=1$.
It is beyond the scope of this article to discuss this in detail, but
let us mention that there exist proposals to generalize
the WGC to multiple $U(1)$ gauge factors\footnote{A generalization to
  include also scalar fields was proposed in \cite{Palti:2017elp}.} \cite{Cheung:2014vva} leading to 
strong constraints also for inflationary models invoking
multiple axion fields \cite{Kim:2004rp,Dimopoulos:2005ac}.

The lesson is that guided by string theory experience and arguments
from black hole physics, one has arrived at a general conjecture
discriminating in simple terms  effective field theories that admit a consistent UV completion (the landscape)
from those that do not (the swampland). In the here derived context one can say
that the WGC conjecture forbids consistent effective field theory models of natural inflation.

More generally, whenever one claims to have found a working 4D effective field model of large (single) field
inflation derived from string theory, one really has to ensure that in particular the
pre-assumed hierarchy of mass scales
\eq{
\label{masshierarchy}
M_{\rm pl}> M_{\rm s}>M_{\rm KK}> M_{\rm mod} >H_{\rm inf} > M_{\Theta}
}
is satisfied. Clearly, for working in an effective 4D field
theory for discussing moduli stabilization, all final masses of the former moduli have to be lighter
than the Kaluza-Klein and the string scale. Moreover, for having a model
of single field inflation the masses of all heavy moduli (excluding
potentially  very light sub-meV axions) have to be larger than the
Hubble scale during inflation. It is important to emphasize that the 
realization of this hierarchy in a final minimum is not yet sufficient, it really
has to hold throughout the entire slow rolling phase.

Before  we discuss the second approach  of realizing axion inflation, namely axion
monodromy, we spend some time on discussing a second conjecture
for discriminating the landscape from the swampland.

\section{The Swampland Distance Conjecture}

In 2006 Ooguri and Vafa had already proposed more conjectures of this type \cite{Ooguri:2006in}.
The most quantitative one has meanwhile been termed the
Swampland Distance Conjecture (SDC).
The inspiration for it comes from a simple dimensional reduction 
of Einstein gravity on a circle. Choosing the ansatz for a 5D metric
\eq{
                       G_{MN} dX^M dX^N= g_{\mu\nu}(x) dx^\mu dx^\nu +
                       r(x)^2 dy^2
}
leads to an effective action for the modulus field $r(x)$
\eq{
\label{actrr}
              S=\int d^4 x \sqrt{g}\,  {1\over (\lambda r)^2} \,\partial_\mu
                r \,\partial^\mu r 
}
where $\lambda$ denotes  a numerical constant. Note that this is the same form of the
metric on moduli space as already encountered in \eqref{axionf}, where however
here it is the kinetic term for a saxionic field $r(x)$ and not an axion.
It followed that the canonically normalized field $\rho(x)$ is
\eq{
\label{logscale}
                    \rho=\lambda^{-1} \log r 
}
which scales logarithmically with $r$.
The mass of Kaluza-Klein modes along the circle are therefore given as
\eq{
                    M_{\rm KK}\sim {n\over r}\sim n\, e^{-\lambda
                      \rho}
}
Thus, for $\rho>\rho_c=\lambda^{-1}$ infinitely many states become
exponentially light indicating a  breakdown of the effective action that has only kept
the massless four-dimensional metric and the radial modulus\footnote{
In \cite{Grimm:2018ohb,Heidenreich:2018kpg} it was shown that integrating
out an infinite tower of exponentially light states is closely related to the appearance
of infinite distances in field space.}.

Motivated by this and similar observations, Ooguri/Vafa proposed that this behavior is a general
property of any effective theory derived from string theory (quantum gravity).
\vspace{0.2cm}

\noindent 
{\it Swampland Distance Conjecture:}

\vspace{0.1cm}
For any point $p_0$ in the continuous scalar moduli space of a
consistent quantum gravity theory, 
there exist other points $p$ at arbitrarily large distance. As the
distance $d(p_0, p)$ diverges,  an infinite tower of states
exponentially light in the distance appears, i.e.
the mass scale of the tower varies as
\eq{
\label{swamp_mass}
m \sim  m_0\, e^{-\lambda d(p_0,p)}\,.
} 
Therefore, beyond the critical distance $d_c=\lambda^{-1}$ the effective field theory
breaks down.

\vspace{0.3cm}
\noindent
Let us make a few comments:

\vspace{-0.2cm}
\begin{itemize}
\item{The distance is measured by the metric on the moduli space and is given by the 
length of the shortest geodesic.}
\item{The conjecture  is about the flat moduli space of a compactification, in which 
axionic fields are compact. Thus, the directions in which arbitrarily large distances appear
involve saxionic directions.}
\item{The original SDC conjecture makes no reference to the value of the critical distance 
$d_c=\lambda^{-1}$ beyond which the effective field theory breaks down.}
\item{Note that the SDC  describes a property of models in the landscape!}
\end{itemize}

If one uses the K\"ahler potential \eqref{kaehlera}, one would find
$\lambda=\sqrt{2/3}M_{\rm pl}$ where we have reintroduced the Planck scale. 
This and similar results for simple models of  axion monodromy 
inflation \cite{Blumenhagen:2015qda, Baume:2016psm} 
led Kl\"awer/Palti  \cite{Klaewer:2016kiy} to go one step further
and conjecture that the critical distance is always of the order
of the natural built in mass scale, namely $\lambda=\alpha M_{\rm pl}$,
where $\alpha$ is a number of order one. 
Adding this  extra piece to the SDC was termed the Refined Swampland Distance Conjecture (RSDC).
Its relation to models of  axion monodromy inflation will be discussed in section 5.

\subsection*{Testing the RSDC on Calabi-Yau moduli spaces}

In \cite{Blumenhagen:2018nts} this RSDC conjecture was challenged by a detailed computation
of distances in the K\"ahler moduli space of Calabi-Yau 
compactifications (see also \cite{Hebecker:2017lxm,Cicoli:2018tcq}).
These K\"ahler moduli spaces do not
only contain (geometric) regions/phases,  where points at infinite distance
exist, but also non-geometric phases, like the
Landau-Ginzburg (LG) phase. These often do only have a finite radius and
therefore do not admit points at infinite distance. To reach the latter one
first has to cross the non-geometric phase and move to another geometric
phase where such a point do exist.

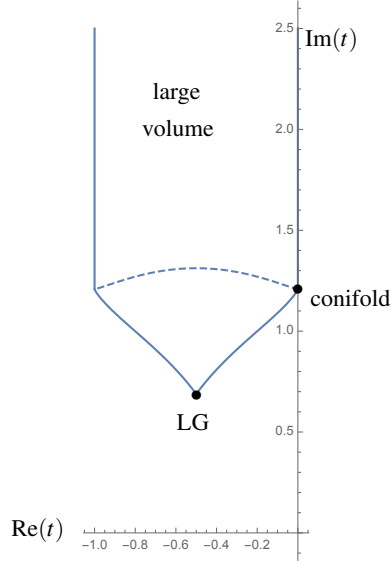
\begin{figure}[ht]
  \centering
  \includegraphics[width=0.2\textwidth]{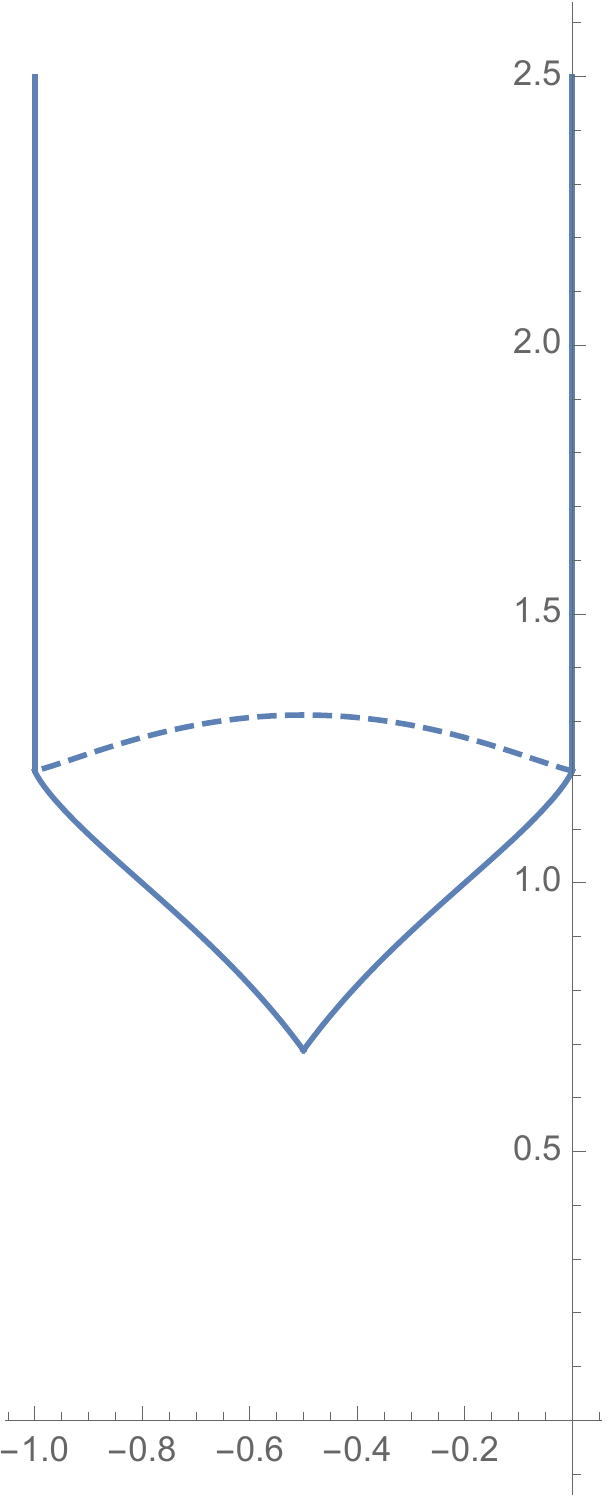}
 \begin{picture}(0,0)
   \put(-115,10){\footnotesize${\rm Re}(t)$}
   \put(-5,195){\footnotesize${\rm Im}(t)$}
    \put(-62,175){\footnotesize large}
     \put(-66,161){\footnotesize volume}
   \put(-53,50){\footnotesize LG}
   \put(-48,61){\footnotesize${\bullet}$}
   \put(-3,97){\footnotesize${\rm conifold}$}
    \put(-10,101){\footnotesize${\bullet}$}
  \end{picture}
   \caption{Sketch of the K\"ahler moduli space of the quintic.}
  \label{fig_quintic}
\end{figure}

\begin{figure}[ht!]
\centering
\begin{tikzpicture}[xscale=0.8,yscale=0.8]
\node[inner sep=0pt] at (0,0)
{\includegraphics[scale=0.8]{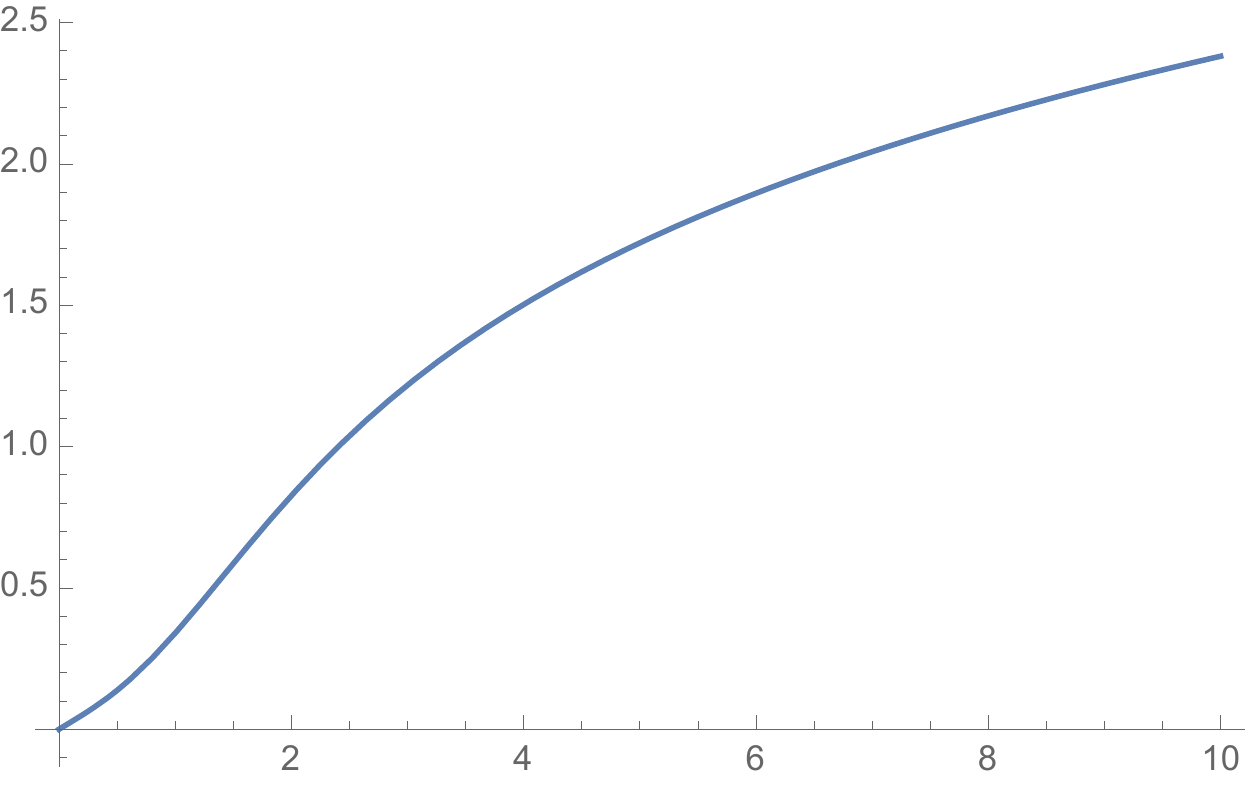}};
\node at (-5.2,4) {$\Theta$};
\node at (6,-4.1) {${\rm Im}\, t$};
\node at (-7.2,-0.5) {$\Theta_0$};
\node at (-7.2,2.3) {$\Theta_c$};
\node at (-2.8,-4.1) {${\rm Im}\, t_0$};
\node at (2,-4.1) {${\rm Im}\, t_c$};
%
\node[align=center,below] at (-1.5,-0.4) {\footnotesize logarithmic\\ \footnotesize behavior\\ \footnotesize relevant};
\node[align=center,below] at (3.2,2.3) {\footnotesize significant\\ \footnotesize decrease of\\ \footnotesize mass scale};
\node[red] at (-2.9,-0.5) {\textbullet};
\node[red] at (1.9,2.25) {\textbullet};
\draw[dashed,red] (-5.7,-0.5) -> (-2.9,-0.5);
\draw[dashed,red] (-2.9,-0.5) -> (-2.9,-3.425);
\draw[dashed,red] (-5.7,2.25) -> (1.9,2.25);
\draw[dashed,red] (1.9,2.25) -> (1.9,-3.4);
\draw [decorate,decoration={brace,amplitude=10pt},xshift=-4pt,yshift=0pt]
(-6.25,-0.5) -- (-6.25,2.25) node [black,midway,xshift=-0.6cm] 
{$\Theta_\lambda$ \ };
\end{tikzpicture}
\caption{\small Expected relation between proper field distance $\Theta$ and ${\rm Im} \, t$.}
\label{nicefigure}
\end{figure}
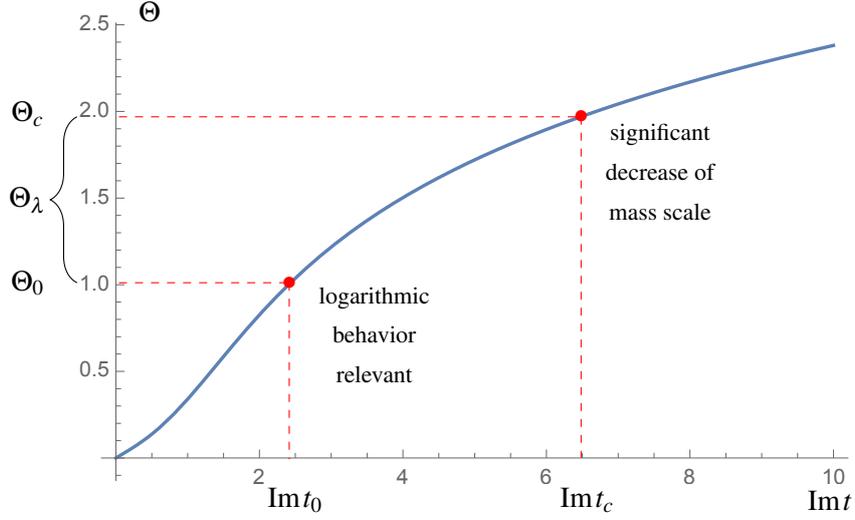

If the distances one can travel along 
geodesics in such  non-geometric phases were  already larger than $M_{\rm pl}$, it would directly
falsify the RSDC \footnote{Similar in spirit, in \cite{Conlon:2016aea}
axion decay constants have been evaluated in the geometric phases for $h^{11}\in \{1,2\}$ CY
threefolds.  Is was found that  they are bounded  from above by 
an order one parameter times the Planck-scale.}.
Let us explain this potential challenge for the concrete example of the K\"ahler moduli
space of the quintic, which looks like as shown in figure \ref{fig_quintic}.
There is only a single complexified K\"ahler modulus $t=B+i J$.
As indicated, there are three distinguished points: the large volume point, the
conifold and the Landau-Ginzburg (LG) point. The LG or Gepner point is the
one of minimal radius. To cover the whole moduli space, one needs at
least two charts, whose radii of convergence are shown by the dashed
arc. Now one can
ask the question whether the RSDC  still
holds for points $p_0$ in the small volume regime. 
Following a geodesic from the LG point to the large volume
regime, one expects that the proper field distance depends on ${\rm Im} \, t$
 like shown in figure \ref{nicefigure}. 
\vspace{0.1cm}

\noindent
As long as one stays in the small volume regime the proper field
distance scales polynomially with ${\rm Im} \, t$ and at some point $({\rm Im} \, t_0,\Theta_0)$ the logarithmic scaling becomes dominant.
As a consequence, we define the critical field distance as the sum
\eq{
               \Theta_c=\Theta_0+\lambda^{-1}\,,
}
which includes the distance $\Theta_0$. 
Clearly, if $\Theta_0$ determined in this way was already much larger
than the Planck-scale, the RSDC  would  be
falsified. Said the other way around, if the RSDC is correct, the
proper field  distance that can be traveled
in the small volume regime must be smaller than $M_{\rm pl}$.

Following this idea, using two different methods to determine
the K\"ahler potential on the  K\"ahler moduli space, in
\cite{Blumenhagen:2018nts} the following two tasks were treated:

\begin{itemize}
\item{Compute the periods and K\"ahler potential in non-geometric
    phases (LG and hybrid) of CY manifolds with $h_{11}\in\{1,2,101\}$}
\item{Compute the critical distances  $\lambda$ and $\Theta_0$ for various geodesics in
    these highly curved moduli spaces\,.}
\end{itemize}

\begin{figure}[h]
  \centering
  \vspace{0.2cm}
 \includegraphics[width=0.60\textwidth]{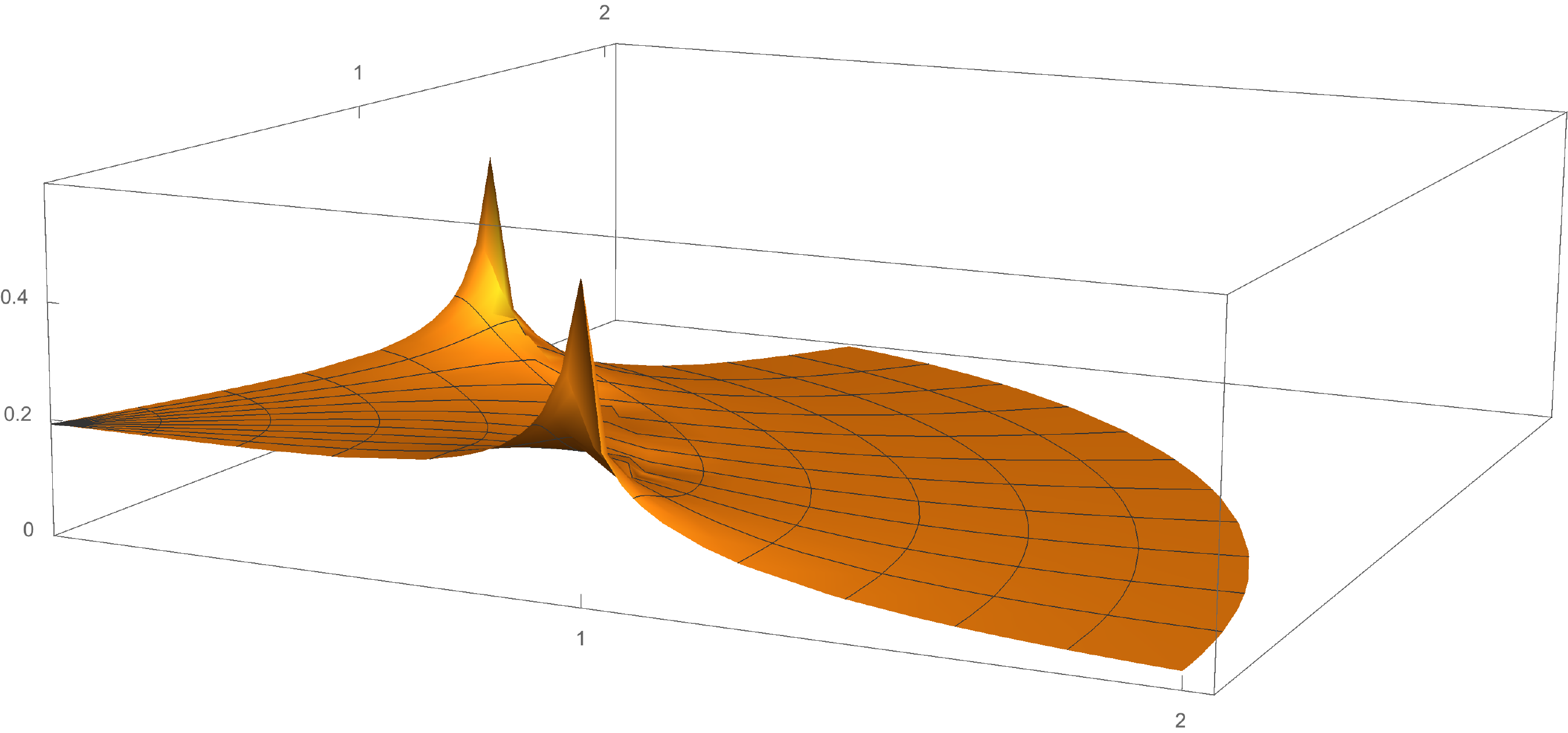}
  \begin{picture}(0,0)
    \put(-35,42){$\psi$}
  \end{picture}
\caption{\small K\"ahler metric for the quintic.}
\label{fig_met}
\end{figure}
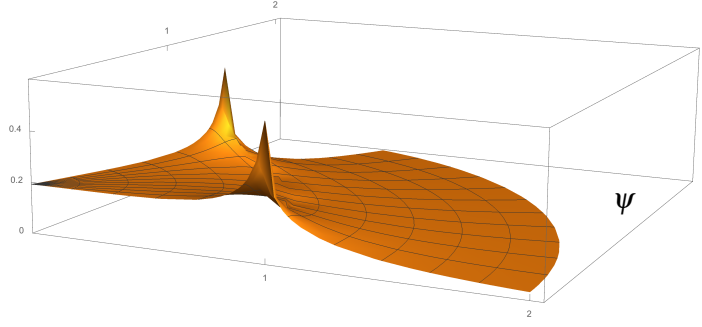

\noindent
For the quintic, the K\"ahler potential in the large volume regime 
takes the familiar form
\eq{
K=- \log \Big( i (t-\ov t)^3  + C + O(e^{-2\pi t})\ldots\Big)
}
whereas in the Landau-Ginzburg phase we obtain ($\psi=|\psi| \exp(i\theta)$ and $|\psi|<1$)
\eq{
   K &=-\log\Big(    \alpha\,|\psi|^{2}
           +\beta \,|\psi|^{4} + \gamma\,|\psi|^{6} 
      + \delta \,|\psi|^{7} \cos(5\theta) +\ldots  \Big)\,.
}
Note that both expressions feature a shift symmetry for the leading order terms,
namely in ${\rm Re}(t)$ and ${\rm Arg}\psi$,
that are broken by higher order corrections. The resulting K\"ahler metric as a function
of $\psi$ is shown in figure \ref{fig_met}. Note that the variable $t$ and $\psi$ are related
via the mirror map.

\noindent
The bisectrix is indeed a geodesic for which  we obtain 
\eq{
               \Theta^{\rm LG}_0=0.43\,,\qquad
               \Theta_\lambda=\lambda^{-1}=\sqrt{\textstyle {3\over 4}} 
}
in accordance with the RSDC. This ist just the simplest example that has been
computed in \cite{Blumenhagen:2018nts}. There, many more tests have been made for other Calabi-Yau threefolds
and more general geodesics. The findings can be summarized by saying that
in all examples one  obtains   $\Theta_0<1$ and  $\lambda^{-1}<1$.
Moreover, studying the recently determined 204 periods of the quintic \cite{Aleshkin:2017fuz}, it was
observed that the distances  in the  LG phase become an order $10^{-2}$ smaller
than in the one-dimensional K\"ahler moduli space of the quintic.
Thus, it is compelling to propose the scaling relation
\eq{
                     \left\langle {\Theta_0\over  {\rm phase}}  \right\rangle \cdot \#({\rm phases})<
                     M_{\rm pl}\, 
}
for the average distance that can be traversed in a non-geometric phase.
Then,  even crossing more non-geometric phases before reaching the large 
volume phase does not help in collecting an overall distance that is much larger
than one. If true, this would be a compelling piece of evidence for  the RSDC.

\section{Axion monodromy inflation}

Let us come back to the question of generating a potential for the initially shift symmetric
axion. As opposed to taking non-perturbative effects into account, that are 
there once the classical string background is specified, one can also
add new classical ingredients.
In the pioneering work \cite{Silverstein:2008sg,McAllister:2008hb} these were D-branes, whose backreaction on the
geometry were inducing a non-trivial potential for the axion (this arose
from the Kalb-Ramond field contribution to the Dirac-Born-Infeld
action $S_{\rm DBI}=\int d^px \sqrt{G+B}\;$).

\subsection*{F-term axion monodromy}

An alternative mechanism was proposed in \cite{Marchesano:2014mla,Blumenhagen:2014gta, Hebecker:2014eua}, namely that the same type of fluxes that one
introduces for moduli stabilization already induces a potential for the axion.
This can be seen from the dimensional reduction of the kinetic terms of the R-R forms
in the ten-dimensional effective supergravity action
\eq{
\label{RRkin}
            S_{10}=\int d^{10}x\, G_p\wedge \star G_p
}
with 
\eq{
              G_p=F_p -H_3\wedge C_{p-3}  + {\cal F}\wedge e^B
}
where $F_p=dC_{p-1}$, $H=dB_2$ and ${\cal F}$ denotes a formal sum of all 
R-R field strengths $F_p$. Note that the fluxes $F_p$ and $H$ are closed and therefore
take values in $H^{p}(M,\mathbb Z)$.  Moreover, the dimensionally reduced
higher gauge  fields $C_p$ and $B$ give rise to axions $\theta$ in four-dimensions. For instance,
in type IIA, the axions $\int_{\Sigma^{(i)}_2} B$ appear in the complexified K\"ahler moduli
$T^{(i)}=\int_{\Sigma^{(i)}_2} B+i \int_{\Sigma^{(i)}_2} J$. 
Since the the gauge fields appear explicitly in $G_p$, the dimensional reduction
of the kinetic terms \eqref{RRkin} directly lead to a scalar potential for axions.

Motivated by the earlier field theory proposals by \cite{Dvali:2005an,Kaloper:2008fb,Kaloper:2011jz},
it was explicitly shown in \cite{Bielleman:2015ina} (see also the review \cite{Valenzuela:2017bvg})
that the resulting scalar potential
takes a very peculiar form, namely it can be expressed in terms
of 4D space-time filling four-forms coupled to the axions
\eq{
          S_{4}=\int d^4 x\, \Big(   - Z_{ab}(\phi) \, F^a_4\wedge \star
            F^b_4+ 2 F_4^a\, \rho(\theta) \Big)+\ldots\,.
}
Here $Z_{ab}(\phi)$ depends only on the saxionic fields $\phi$ and $\rho(\theta)$ depends
on the axionic fields and the fluxes turned on. Integrating out the 4D three-forms in $F^a_4=dC^a$
and choosing the integration constant to be vanishing, one arrives at the scalar potential
\eq{
            V=  \big(Z^{-1}(\phi)\big)^{ab} \, \rho_a(\theta) \,\rho_b(\theta)\,,
}
that nicely separates the dependence on the saxions and the axions.
To illustrate what is going here, let us consider a simplified field theory model,
where  the Lagrangian of a single  axion  $\theta$ coupled to a
three-form gauge field reads
\eq{
            {\cal L}= - f^2 d\theta\wedge \star d\theta -F_4\wedge \star
            F_4+ 2 F_4( m\theta +f_0)\,.
}
As we have seen, in string theory  $f_0$ and $m$ can be thought of as quantized background fluxes.
There, the axion decay constant $f$ will also be (saxionic) moduli dependent.
The resulting equation of motion for $C_3$ takes the simple form
\eq{
               d\star F_4=d(m\theta+f_0)\quad\Rightarrow \quad  \star F_4=f_0+m\theta
}
where the integration constant was set to zero. Introducing this back into the action
 leads to an effective potential 
\eq{
                  V=(f_0+m\theta)^2\,.
}
We observe that the scalar potential and $F_4$ are  invariant under the extended shift symmetry
\eq{
          \theta\to \theta - c/m\, \qquad  f_0\to f_0 +c \,.
}
Therefore, the system still preserves the shift symmetry, that is broken
spontaneously by a choice of branch $f_0,m$.
Moreover, the shift symmetry and the gauge symmetry of $C_3$ strongly
constrain higher order corrections.
They must be functions of $F_4$, i.e.
\eq{
               \delta V\sim \sum (F_4)^{2n} \sim \sum \left
                 ({V_0}\right)^n 
}
so that even in the trans-Planckian regime  $\delta \theta\gg 1$, 
as long as $\delta V\ll 1$ one controls the expansion.
In other words, this way of breaking the shift symmetry (on a branch) is not generic
but of a special kind that still allows to control the Planck suppressed operators.

\subsection*{Inflation}

Using these flux induced potentials for an  axion to realize inflation goes under the name
of axion monodromy inflation. On a fixed branch the scalar potential increases
by a certain amount when one traverses over one period of the former
shift symmetric axion. Compared to the models of natural inflation
the range of the axion $\theta$ is now disentangled from the axion decay constant
(depending on the saxions) so that the WGC conjecture does not 
 provide any direct constraint on the field range of the axion.

The question is whether there exist other
problems of controlling the effective field theory along a trajectory that involves
trans-Planckian field ranges for the axion. First, one realizes that 
 in full examples the issue of axion inflation cannot be disentangled from
the issue of moduli stabilization. It is the same flux induced potential
that is responsible for both. Therefore, one has to make sure that the 
fluxes can be turned on in such a way, as to allow that the lightest (heavy)
modulus is an axion (see e.g. \cite{Blumenhagen:2014nba}). This would then be the candidate for the inflaton.
Since the axion/inflaton receives its small mass from a tree-level effect, 
it is unnatural to expect that moduli getting their mass from a non-perturbative
effect come out more massive than the inflaton. In other words,
{\it all} other heavy moduli should better also be stabilized by tree-level fluxes.

Let us assume that using a 4D low-energy effective action, one has stabilized all moduli, 
so that a canonically normalized  axion $\Theta$ is the lightest one 
with mass $M_{\Theta}$ and  all the remaining moduli  have a higher
mass  $M_{\Theta}\ll M_{\rm heavy}\ll M_{\rm KK}$.
For inflation, the axion gets displaced from its minimum and, as mentioned
already before, along the entire  slow rolling phase  the hierarchy \eqref{masshierarchy} must be 
intact.  However, when $\theta$ is displaced from its minimum, generically the minima
of the saxions will  also change,
\eq{
\label{saxion}
r(\theta)=r_0+\delta r(\theta)
}
where $r_0$ denotes the vacuum expectation value of the scalar $r$
at the minimum of the potential, i.e. when $\theta$ is also at its minimum. 
By plugging this back into the effective theory, the scalar potential and the kinetic
term for the inflaton can be substantially modified. In other words,
the inflationary trajectory is no longer only along $\theta$ but
corresponds to a combination of $\theta$ and $r$. It has been pointed out in 
\cite{Dong:2010in} that this backreaction
leads to a flattening of the inflaton potential.

Studying concrete examples, in \cite{Blumenhagen:2015qda,Baume:2016psm} it was realized  that
the displacement of the saxions will generically backreact on the
kinetic metric of the inflaton leading at best to a logarithmic
behavior of the proper field distance at large field. More concretely,
\eq{
\label{hannover96}
\Theta=\int \sqrt{K_{\theta\theta}(s)}\; d\theta\sim \int \frac{1}{r(\theta)}\sim \frac{1}{\lambda} \log(\theta)
}
where we have used that $K\sim -\log(r)$ with $r$ being the saxionic
partner of the inflaton, and that for large field excursions $\delta
r(\theta)\simeq \lambda \theta$ (which was found for concrete examples). 
In \eqref{hannover96}, $\Theta$ is
the canonically normalized inflaton field. 
This behavior is reminiscent of the logarithmic scaling in the SDC \eqref{logscale},
with the difference that now we  are dealing with axions that are not flat directions
but are moving in a potential.
As in the SDC,  the mass of Kaluza-Klein modes scales exponentially with the 
field excursion of the axion
\eq{
                    M_{\rm KK}\sim {n\over r}\sim {n\over \lambda}\, e^{-\lambda
                      \rho}\,.
}
Whether  trans-Planckian field ranges in $\Theta$ are under control
now depends on the value of $\lambda$. In concrete  examples this is related
to the ratio of light to heavy moduli masses.
\eq{
              \lambda=\bigg({M_\Theta\over M_{\rm heavy}}\bigg)^p\,
}
where values  $p=1/2,1$ have been found in concrete models.
Therefore a mass hierarchy between the inflaton and
the saxionic  moduli can help to delay the backreaction effects which are not
anymore tied to the Planck mass in an obvious way. 
To have parametrical control over this mass hierarchy, requires the minimum of the
potential to satisfy the following condition \cite{Blumenhagen:2014nba, Valenzuela:2016yny}: 
$\Theta_c=\lambda^{-1}$ will be tuneable if one can parametrically set the inflaton mass to zero without 
destabilizing the other scalars. In order to see whether this is possible
in (quasi-) realistic set-ups, we consider a simple example from \cite{Blumenhagen:2017cxt}, that realizes
many of the ingredients that we need.

\subsubsection*{An example with a D-brane modulus}

We consider the  so-called type IIB $STU$-model extended by a complex open string
modulus $\Phi$ that parametrizes the transversal deformation of a
D7-brane. Such a model has also been studied in \cite{Bielleman:2016olv}. 
We work in the framework of effective four-dimensional ${\rm N}=1$ supergravity,
where all string and Kaluza-Klein modes have been integrated out.
The four complexified  moduli are
\eq{
        S=c+is\,,\qquad T=\rho+i\tau\,, \qquad U=v+iu\,,\qquad
        \Phi=\theta+i\varphi\,
} 
where the real  parts are axion-like scalars. At large values of the saxions $(s,\tau,u)$, 
the K\"ahler potential is given at leading order  as
\eq{
\label{kaehleropen}
          K=-3\log(-i(T-\ov T))&-2\log (-i(U-\ov U)) 
      -\log\left[ -(S-\ov S)(U-\ov U) +{\textstyle {1\over 2}}
            (\Phi-\ov \Phi)^2\right]\,.
}
Now we turn on fluxes to generate the superpotential
\eq{
            W={\mathfrak f}_0-3{\mathfrak f}_2\, U^2 +h\, S\,U+q\, T\, U+\mu\, \Phi^2 \, 
}
where the fluxes are considered to be integers.  This model is of the
flux-scaling type  discussed in \cite{Blumenhagen:2015kja,Blumenhagen:2015xpa}.
In type IIB the term involving the K\"ahler modulus
$T$ is generated by a so-called non-geometric Q-flux.
This flux breaks the no-scale structure that is present for $q=0$ and which was the starting
point for the KKLT and the LVS scenario of K\"ahler moduli stabilization via non-perturbative effects.
As mentioned, as we want to identify the real part of $\Phi$ with  the inflation, stabilizing the K\"ahler modulus
in the latter non-perturbative way would lead to the wrong mass hierarchy.

The K\"ahler- and the superpotential  data specify the supergravity model and one can compute the scalar potential.
It admits an analytically solvable  non-supersymmetric tachyon-free AdS minimum at
\eq{    s_0&={2^{7\over 4}\cdot 3^{1\over 2}\over 5^{1\over 4}} {({\mathfrak f}_0\,
    {\mathfrak f}_2)^{1\over 2}\over h}\,,\qquad 
    \tau_0={5^{3\over 4}\cdot 3^{1\over 2}\over 2^{1\over 4}} {({\mathfrak f}_0\,
    {\mathfrak f}_2)^{1\over 2}\over q}\,,\qquad 
    u_0={1\over 10^{1\over 4} \cdot 3^{1\over 2}} \left({{\mathfrak f}_0\over
    {\mathfrak f}_2}\right)^{1\over 2}\\[0.1cm]  \varphi_0&=0 \,,\qquad
   v_0=hc_0+q\rho_0=\theta_0=0 \, ,
}
leaving one axionic direction unconstrained.   The value of the scalar
potential in the AdS minimum is
\eq{
\label{vacuumenergy}
                 V_0=-{1\over 120\cdot 3^{1\over 2}\cdot 10^{1\over
                     4}}{h\,q^3 \over {\mathfrak f}_0^{3\over 2}\, {\mathfrak f}_2^{1\over 2}}\,.
}

For realizing inflation, one 
now needs to make the assumption that this minimum can be uplifted to Minkowski, i.e. $V_0=0$,
without destabilizing the moduli and only slightly changing their vacuum expectation values.
For recent critical discussions on the existence of bona-fide de Sitter vacua in string theory 
and the consistency of up-lift mechanisms please consult \cite{Sethi:2017phn,Danielsson:2018ztv}.  
To this discussion one can add another argument supporting the claim
that the correct description  of an uplift by an $\ov{D3}$-brane
placed in a highly warped throat is not yet fully understood.
In most cases, one simply adds to the supergravity generated
scalar potential an uplift term
\eq{
\label{uplift}
   V=V_{\rm SUGRA}+V_{\rm up}=V_{\rm SUGRA}+{\epsilon\over \tau^2}
}
with $\epsilon\ll 1$ in the 
strongly  warped region. However, the complex structure modulus $Z$ governing
how close one is to a conifold singularity also needs to be stabilized
before. 
It was shown in \cite{Blumenhagen:2016bfp}
that the self-consistency of the used supergravity action requires that the physical volume
of the corresponding 3-cycle $A$ must still be larger than one, i.e.
\eq{
                        {\rm Vol} (A)=   {\cal V}^{1\over 2}    \int_A \Omega_3 = (|Z|^2\, {\cal V})^{1\over 2}>1
}
so that one stays  in the dilute flux limit. Here $ {\cal V}$ is the
total volume of the CY threefold.
This means that the strongly warped region is outside the validity of the used 
supergravity  effective action.
Therefore, the ansatz \eqref{uplift} for the total scalar potential is highly questionable.
We note that  for the upcoming discussion of the backreaction, the existence of an uplift is not 
really relevant.

Now coming back to our example, 
the mass-matrix has  the eigenvalues\footnote{Introducing back factors
  of $M_{\rm pl}$ we note that $|V_0|\sim M^2_{\rm heavy}\, M^2_{\rm pl}$.}
\eq{
\label{massesa}
         M^2_{\rm closed}=\nu_i\, {h\, q^3\over {\mathfrak f}_0^{3\over 2}\,
           {\mathfrak f}_2^{1\over 2}}
}
with $\nu\in\{0,0.0001,0.0019,0.0029,0.0117,0.0162\}$ and 
\eq{
\label{massesb}
M^2_{\phi}& \simeq 0.002\, {h\,q^3\over  {\mathfrak f}_0^{3\over 2}\,
           {\mathfrak f}_2^{1\over 2}}\,,\qquad
M^2_{\theta}\simeq 0.021\, {\mu\,q^3\over  {\mathfrak f}_0^{3\over 2}\,
           {\mathfrak f}_2^{1\over 2}}\,.
}
Therefore,  the open string axion $\Theta$ can be  {\it parametrically} lighter than all
the other massive moduli, indeed 
\eq{
     {M_{\rm heavy}\over M_{\Theta}}\sim \sqrt{h\over \mu}=\lambda^{-1}\,.
}
From the former discussion, one expects
that $\lambda=\sqrt{\mu/ h}$ is the  flux dependent parameter 
that controls the backreaction of the inflaton onto the other moduli.
 Let us analyze this in more detail under the assumption
$\lambda\ll 1$. Up to subleading corrections of order ${\cal O}(\lambda^{2})$, 
the conditions for the backreacted minima can be solved 
\eq{  
\label{back_vevs}
s_0(\theta)&\sim {2^{7\over 4}\, 3^{1\over 2}\over 5^{1\over 4}}
  {({\mathfrak f}_0+\mu\theta^2)^{1\over 2}\,
    {\mathfrak f}_2^{1\over 2}\over h}\,,\qquad
            \tau_0(\theta)\sim {5^{3\over 4}\, 3^{1\over 2}\over 2^{1\over 4}} {({\mathfrak f}_0+\mu\theta^2)^{1\over 2}\,
    {\mathfrak f}_2^{1\over 2}\over q}\\
      u_0(\theta)&\sim {1\over 10^{1\over 4} \, 3^{1\over 2}} \left({{\mathfrak f}_0+\mu\theta^2\over
    {\mathfrak f}_2}\right)^{1\over 2}\,
}
with all other fields sitting in their minimum at zero.
Thus, the critical value of $\theta$ where the backreaction becomes
significant is  $\theta_{\rm c}=\sqrt{f_0\over\mu}$.
The kinetic term for the inflaton becomes
\eq{
      {\cal L}^{\rm ax}_{\rm kin}=K_{\Phi\ov\Phi} \,\partial_\mu
      \theta \partial^\mu \theta={1\over 8}\sqrt{5\over 2}\, {h\over f_0+\mu\theta^2}
      \left({\partial \theta}\right)^2
}
so that the critical value for the canonically normalized inflaton
field $\Theta$ is 
\eq{
       \Theta_{\rm c}=\gamma
       \sqrt{h\over f_0}\theta_{\rm c}=\gamma \sqrt{h\over \mu}=\gamma \lambda^{-1}\,
}
with $\gamma={1\over 2} \left({5\over 2}\right)^{1\over 4}=0.63$.
Therefore, from this perspective, for $\lambda\ll 1$ and $\Theta\ll
\Theta_{\rm c}$ the backreaction
can be neglected and one gets a polynomial  effective potential for the inflaton
(after adding a  constant uplift).
Thus, it seems that by parametrically choosing $\Theta_{\rm c}\sim
\lambda^{-1}> 10$ one can achieve a stringy model featuring large field
inflation\footnote{This is consistent with the
observation already made in \cite{Bielleman:2016olv} for a more complicated, only
numerically treatable open string model (without non-geometric fluxes).}.

Beyond the critical value,  the kinetic term for the inflaton takes
the form
\eq{
      {\cal L}^{\rm ax}_{\rm kin}={1\over 8}\sqrt{5\over 2}\, {h\over \mu}
      \left({\partial \theta\over \theta}\right)^2
}
so that the canonically normalized inflaton shows the logarithmic  behavior
\eq{
      \Theta=\Theta_{\rm c}\, \log\left({\theta\over \theta_{\rm c}}\right)\simeq \frac{1}{\lambda}\,\log\theta \simeq {M_{\rm heavy}\over M_{\Theta}}\,\log\theta\,.
}
This  is familiar from the RSDC. The difference is that here it is an axionic field
that can go to large distances and that the parameter $\lambda^{-1}\sim  \sqrt{h\over \mu}$
can be parametrically larger than one so that trans-Planckian distances seem to be controlled.
The reason behind it is that for $\lambda^{-1}\gg 1$ the inflationary trajectory 
goes mostly in the axionic direction with only
a slight movement in the (dangerous) saxionic directions


However, we have only analyzed one aspect of the quantum gravity embedding of trans-Planckian
axionic field directions. Since we have more mass scales in the problem like the hierarchy
between the heavy moduli masses and the Kaluza-Klein mass scale, other issues might occur.
To see what happens let us compute  the various mass scales, like
string scale, Kaluza-Klein scales, heavy moduli masses and the inflaton
mass. 
Since we are heading for systematic effects, we  will not be concerned with model dependent numerical prefactors,
but will focus on desired  mass hierarchies that are guaranteed or  spoiled
parametrically. The relevant masses scale in the
following way with the fluxes (recall that we set $M_{\rm pl}=1$):
The string scale is
\eq{
\label{stringscalemin}
                      M^2_{\rm s}\sim { 1\over \tau^{3\over
                          2}\, s^{1\over 2}}\sim {h^{1\over 2}\,
                          q^{3\over 2}\over {\mathfrak f}_0\, {\mathfrak f}_2} \,.
}
Moreover, considering our model as being realized on  the isotropic
$T^6$, we now have {\it two} Kaluza-Klein scales for $u>1$, yielding
a {\it heavy} and a {\it light} Kaluza-Klein mass
\eq{
                      M^2_{\rm KK}\sim { 1\over \tau^{2} }\,
                      u^{\pm  1} \quad\Longrightarrow\quad 
             M^2_{\rm KK,h}\sim {q^2\over {\mathfrak f}_0^{1\over 2}\, {\mathfrak f}_2^{3\over 2}} \,,\qquad
           M^2_{\rm KK,l}\sim {q^2\over {\mathfrak f}_0^{3\over 2}\, {\mathfrak f}_2^{1\over
               2}} \,.
}
Recall that the mass of the heavy moduli and the inflaton scaled as
\eq{
         M^2_{\rm heavy}\sim  {h\, q^3\over {\mathfrak f}_0^{3\over 2}\,
           {\mathfrak f}_2^{1\over 2}}\,,\qquad
M^2_{\Theta}\sim  {\mu\, q^3\over {\mathfrak f}_0^{3\over 2}\,
           {\mathfrak f}_2^{1\over 2}}\,.
}
We can also evaluate the various
mass-scales in the large field regime. Due to \eqref{back_vevs}, this
means that we just have to change  
\eq{
{\mathfrak f}_0\to {\mathfrak f}_0\left({\theta\over \theta_{\rm c}}\right)^2\to
{\mathfrak f}_0\exp\left(2{\Theta\over \Theta_{\rm c}}\right)
}
so that  the  various mass scales become
\eq{
                      M^2_{i}=  M^2_{i}\big\vert_0\,
         \exp\left(-\mu_i {\Theta\over \Theta_{\rm c}}\right)\,.
}
with the integer coefficient $\mu_s=2, \mu_{{\rm KK,h}}=1, \mu_{{\rm
    KK,l}}=\mu_{\rm heavy}=\mu_{\Theta}=3$.
Note that all these mass scales become exponentially light for large $\Theta$.
We observe that  the quotient of the  {\it light} KK-mass and the heavy moduli 
mass is constant along the trajectory\footnote{For the flux
  supported AdS vacuum, the curvature radius of the AdS space is given
  by $L_{\rm AdS}\sim M_{\rm pl}/V_0^{1\over 2}\sim M^{-1}_{\rm
    heavy}$. Therefore for the ratio of the length scales $L_{\rm AdS}/L_{\rm KK,l}\sim M_{\rm
    KK,l}/M_{\rm heavy}$ one finds the same ratio of mass scales as in
  \eqref{ratio_essentielc}. This means that parametrically the initial
AdS minimum does not allow a clear separation of the length scales  of the
compact and the non-compact space.}
\eq{    
\label{ratio_essentielc}
{M^2_{\rm KK,l}\over M^2_{\rm mod}}\sim {1\over h\, q} \,.
}
Now, for quantized fluxes, one can only tune  $\lambda^{-1}\sim  \sqrt{h\over \mu}$
large  by choosing a large flux parameter $h\gg \mu$. However, in this regime parametrically
one gets $M^2_{\rm KK,l}\ll M^2_{\rm mod}$ so that the light KK modes become lighter than 
the moduli masses, thus spoiling the use of the low energy effective supergravity action.
Note that we are not talking here about model dependent numerical coefficient, but  about the
parametrical dependence and this goes in the wrong direction and makes
trans-Planckian field ranges of $\Theta$ at least unnatural.

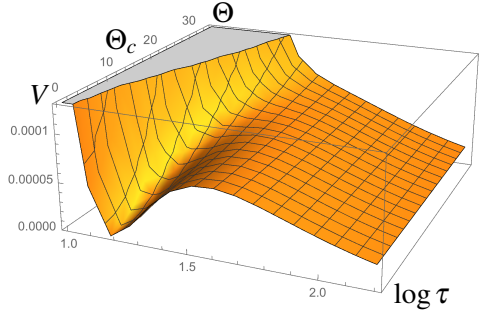
\begin{figure}[ht]
  \centering
   \includegraphics[width=0.41\textwidth]{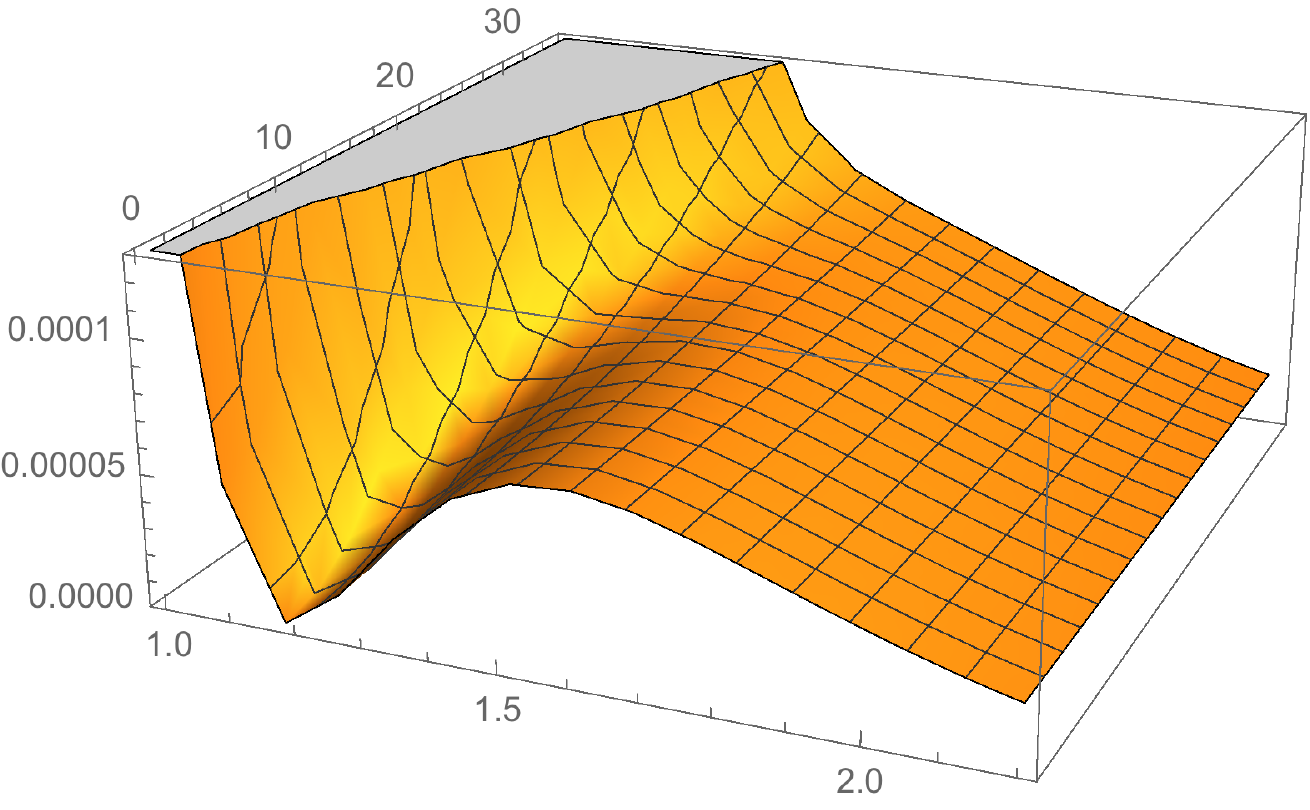}
\begin{picture}(0,0)
   \put(-35,0){$\log \tau$}
   \put(-171,75){$V$}
    \put(-103,106){$\Theta$} 
    \put(-143,96){$\Theta_c$} 
  \end{picture}
 \caption{A backreacted trajectory that destabilizes  at $\Theta_c$. $\tau$
   denotes a saxionic field.}
  \label{fig_valley}
\end{figure}

Thus, in this concrete (admittedly not completely perfect) 
string motivated model, we found evidence that the distance in 
proper field space $\Theta$, where the
logarithmic behavior sets in, is around the Planck-scale and
cannot be much increased without invalidating the effective theory. 
Of course this is just a single example but  a similar behavior has been
found in all of the models studied in
\cite{Blumenhagen:2017cxt,Blumenhagen:2015kja}. There, e.g. for the
KKLT and LVS type models  from \cite{Buchmuller:2015oma},
another issue was observed
namely that for the anti D3-brane uplifted models the backreacted valley along which inflation is supposed to happen
destabilizes at around the scale $\Theta_c$. This behavior is shown
in figure \ref{fig_valley}.

\noindent
All these findings point towards the  axionic extension of the RSDC  that says:
{\it  Even for axion monodromy models
with potentially infinite distances in axionic directions, the validity of the effective field
theory breaks down for field excursions of the order of the Planck-scale.} 

\begin{figure}[ht]
  \centering
  \vspace{0.4cm}
  \includegraphics[width=0.66\textwidth]{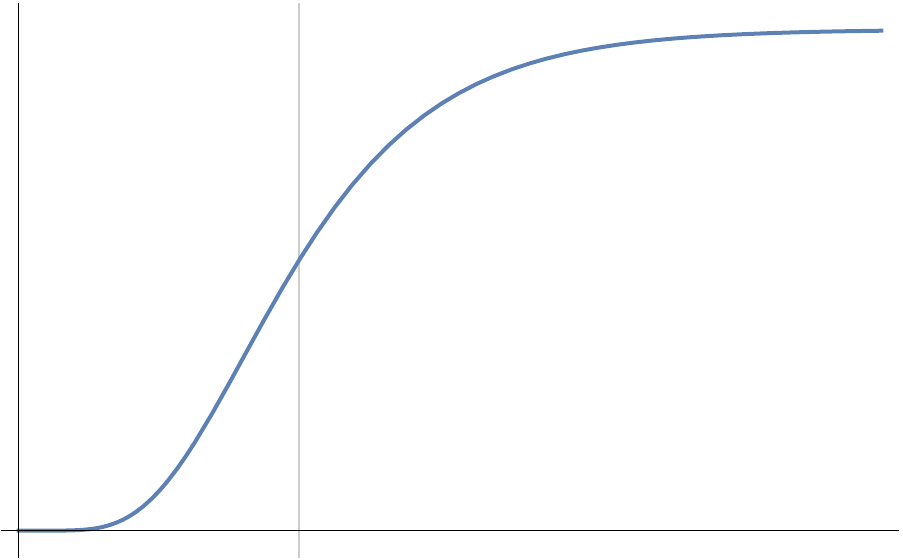}
  \begin{picture}(0,0)
    \put(0,5){$\Theta$}
    \put(-197,-5){$\Theta_c$}
    \put(-285,180){$V(\Theta)$}
    \put(-277,60){${\rm Polynomial}$}
     \put(-277,45){${\rm inflation}$}
      \put(-270,150){\bf sub Planckian}
     \put(-270,135){\bf due to RSDC}
    \put(-170,170){Starobinsky-like inflation}
    \put(-170,60){\bf Invalidity of EFT}
    \put(-170,45){\bf due to SDC}
  \end{picture}
\caption{Large field inflation EFTs derived from UV completion.}
  \label{fig_lfinfl}
\end{figure}
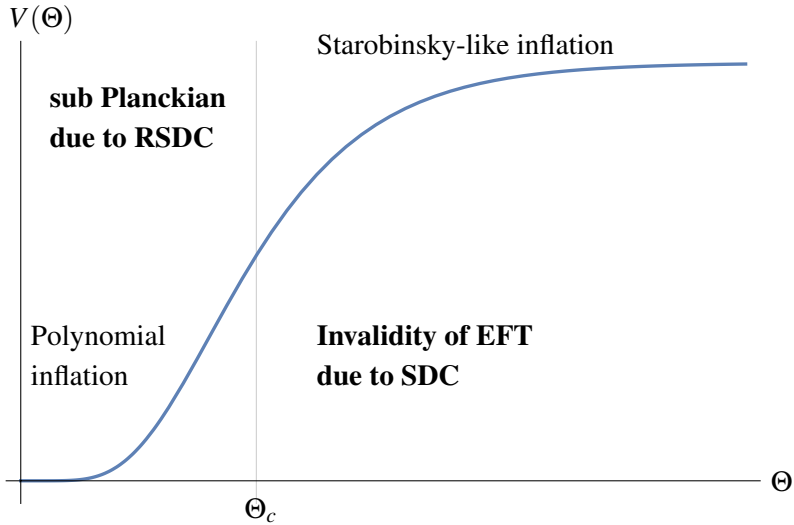

As we have seen, the
validity of this conjecture is closely related to the possibility
to  stabilize the moduli such that a
single axion is parametrically lighter than all the other moduli.
Note that here one is not dealing with geodesic distances in a moduli space,
but with actual trajectories in a field space with a scalar potential.
Thus, one has two competing effects and the true trajectory will
be somewhere between a geodesic and a valley in  the potential.
Since the axionic RSDC conjecture is a very general and very strong statement, more
evidence is needed to support it or to disprove it by a convincing counter-example.
In the moment  we can say that so far no
fully compelling string model of axion monodromy inflation has
been worked out that admits controllable trans-Planckian field ranges.
If the axionic RSDC is correct one gets a picture like shown in the figure \ref{fig_lfinfl}.

\noindent
Therefore, all these developments could culminate in a no-go theorem of the type:
In string theory (quantum gravity) it is impossible to achieve a
{\it parametrically} controllable effective field theory model of large (single)
field inflation. The tensor-to-scalar ratio is thus bounded
from above by $r\lessapprox 10^{-3}$. If true, it offers an experimentally
testable way to falsify string theory. However, it could also be that
the highly non-trivial axionic extension of the RSDC is biased by the examples one has
studied so far. More work is needed to settle this important question.

\section{Comments on quintessence}

As we mentioned in section 2, the mass scales appearing in
quintessence models   are
extremely tiny 
\eq{
          H\sim M_{\Theta}\sim 10^{-33}\, {\rm eV}\,,\qquad
          V_{\rm quint}\sim (10^{-3}\,{\rm eV})^4
}
and therefore Planck-suppressed operators are even more  dangerous to
destroy the  control over the slow-roll regime. 
If the quintessence field is a saxion  then there is no mechanisms
to control the dangerous Planck suppressed operators and 
during the rolling some of the coupling constants could
be time dependent. Moreover, there is the general concern of a long
range fifth force.

Thus, again one is  led to axionic fields with  an initial shift symmetry to realize
quintessence (see \cite{Kaloper:2008qs,Panda:2010uq}). For such models
with a potential $V(\Theta)\sim \Theta^p$ the slow-roll
conditions are met for $\Theta\gg p M_{\rm pl}/\sqrt 2$. From
the measured value  of the parameter in the equation of state
$w=-1.006\pm 0.045$, one can infer that
$\,2.8 p\,  M_{\rm pl}\lessapprox \Theta < \infty$. 
Moreover, the number of e-foldings up to the present time is of order
one, as  $N_e\sim H t_0\sim 1$. Therefore, depending on the values of
$w$ and $p$, one could stay closer to Planck-size field excursion than
for large field inflation\footnote{I thank Nemanja  Kaloper for  pointing
  this out.}.

For models with a non-perturbatively generated periodic potential, one faces
the same problem as for inflation. In the trans-Planckian regime one
still needs $f>1$  and a very small $\exp(-S_{\rm inst})$ factor,
which is not consistent with the WGC. For models of axion
monodromy, even though the distance the field has to roll is smaller
than for inflation,   one still  has to work in the (close to) trans-Planckian regime.
The main challenge though is that in a full string theory setup one needs to stabilize the moduli such that the
mass of the axionic quintessence field is by many orders of magnitude smaller
than the heavy saxionic moduli masses. Due to the cosmological moduli problem
these must satisfy $M_{\rm heavy}>30\,$ TeV so that
\eq{
                       \lambda\sim {M_{\Theta}\over M_{\rm heavy}}  \sim 10^{-45}\,.
}                        
For the previously discussed concrete example this involves huge fluxes that are 
certainly not consistent with tadpole cancellation and the hierarchy between
the moduli and the KK-masses. In more general words, the axionic RSDC will also 
impose very strong constraints  for axionic models of quintessence to be 
realizable in effective theories derived from  string
theory.
Therefore,  effective axionic quintessence models might  be in the
swampland, 
as well.

\section{Outlook}

In the moment, we are witnessing a paradigm shift  in the development of string theory or better string
phenomenology where people take the message from string model building
obstacles seriously and make an effort to extract general conceptual rules that govern any 
theory of quantum gravity. Here, besides general arguments based on
black holes, string theory provides an important source of 
insight as it is  a theory of quantum gravity that methodologically and technically is quite well
understood. However, behind all its technical ingredients it seems to hide some basic
principles of quantum gravity.  
As we have reviewed, once these more general principles are revealed,
they could potentially lead to no-go theorems that in principle
allow one to experimentally falsify string theory.

In this article we were mostly concerned  with a UV completion of models of large field inflation with
a measurable tensor-to-scalar ratio. The arguments follow the logic that 
the interesting parametric region for realizing it  lies out of control of the effective action
one is using. This is not a full no-go theorem, as it might only indicate that 
one is using an unsuitable framework and should better use the full string theory without
referring to an effective action. Of course, how this can be done in
practice is unexplored.

Finally, let us also mention two more conjectures discriminating properties of the
string landscape from those of the swampland. For more details we refer to the original
literature.
In \cite{Ooguri:2016pdq} it has been proposed that the equal sign in the WGC is only for BPS states
and that as a consequence every (flux supported) non-supersymmetric
AdS vacuum is (non-perturbatively) unstable.  This has immediate consequences
for  non-supersymmetric AdS-CFT duality, as on the field theory side a
finite life-time sets a scale and therefore  violates the  conformal symmetry.

Maybe even more drastic, in \cite{Brennan:2017rbf} a potential conjecture was mentioned that claims
that meta-stable de Sitter vacua are also in the swampland of string theory. 
This was also argued for in \cite{Danielsson:2018ztv} where a critical look on existing attempts to construct
dS vacua was given.
Such a no-go theorem would imply that our universe cannot be in a meta-stable state
but must have some still rolling scalar fields. This is reminiscent of models
of quintessence. As we have just mentioned, if the conjectured axionic RSDC indeed
holds, string derived effective models of this type
will be   plagued with the same kind of control issues  as large field
inflationary models\footnote{As an analogy let us mention that 
 by analyzing the question of the (non-)existence of an S-matrix \cite{Hellerman:2001yi},
it was found that rolling eternal quintessence models  behave
analogously to  de Sitter vacua}. Admittedly, the axionic RSDC is not fully
established yet and is motivated by considering a few examples. More
work is needed to see whether our intuition is maybe misguided by the
lamppost we were looking under.

\vspace{1cm}

\noindent
\emph{Acknowledgments:} It is pleasure to thank Cesar Damian, Anamaria Font, Michael
Fuchs, Daniela Herschmann, Daniel Kl\"awer,  Erik Plauschinn, Lorenz
Schlechter, Yuta Sekiguchi, Rui Sun, Irene Valenzuela
and  Florian Wolf for collaboration on the topics reviewed in this article.
I also thank Eran Palti for enlightening discussions.



\end{document}